\documentclass[3p,times]{elsarticle}
\usepackage{ecrc}
\usepackage{epstopdf}
\usepackage{graphicx}
\usepackage{amsmath,amssymb}
\usepackage{subfigure}
\usepackage{color}
\usepackage{soul}

\volume{00}
\firstpage{1}
\journalname{Nuclear Physics A}
\runauth{I.~Kozlov et al.}
\jid{nupha}
\jnltitlelogo{Nuclear Physics A}

\begin{document}

\begin{frontmatter}

\title{Signatures of collective behavior in small systems}

\author[label1]{I. Kozlov}
\ead{ikozlov@physics.mcgill.ca}
\author[label1,label2]{M. Luzum}
%\ead{mluzum@physics.mcgill.ca}
\author[label1]{G.~S. Denicol}
\author[label1]{S. Jeon}
\author[label1,label3]{Gale C.}
\address[label1]{
McGill University, Department of Physics, 3600 University Street, Montreal QC H3A 2T8, Canada}
\address[label2]{
Lawrence Berkeley National Laboratory, Berkeley, CA 94720, USA}
\address[label3]{
Frankfurt Institute for Advanced Studies, Ruth-Moufang-Str. 1, D-60438 Frankfurt am Main, Germany}

\begin{abstract}
We perform 3+1D viscous hydrodynamics calculations of proton-nucleus (pA)
and nucleus-nucleus (AA) collisions. Our goal is to understand the apparent collective behavior
recently observed in pA collisions and to verify whether the highest multiplicity
collision systems can be accurately described as a relativistic fluid. 
We compare our calculations of flow variables to existing measurements, 
and demonstrate that hydrodynamics correctly captures the measured trends. 
We show that our predictions for  the pair correlation observable $r_n$ are validated by recent experimental pA measurements, 
and that our results are sensitive to the granularity of the initial state. 
We also compare our results with measurements done for nucleus-nucleus collisions. 
\end{abstract}


\end{frontmatter}

\section{Introduction}

\label{intro}
It is currently believed that ultrarelativistic heavy-ion collisions at the Relativistic Heavy-Ion Collider (RHIC) and at the Large Hadron Collider
(LHC) are able to reach temperatures high enough to create and study the quark-gluon plasma (QGP) in a controlled experimental environment. One of the most surprising results obtained at RHIC, and more recently, at the LHC, is that this novel state of nuclear matter behaves as an almost prefect fluid, with one of the smallest shear viscosity-to-entropy density ratios in nature.

The main point of colliding heavy ions instead of, e.g. protons, was always to create a system that is large enough to achieve or approach thermodynamic equilibrium. Only if this is accomplished it becomes possible to study the thermodynamic and transport properties of the bulk nuclear matter. Recently, some of the same signs of collective behavior initially observed in heavy-ion collisions more than a decade ago at RHIC, were also seen in high multiplicity p-Pb collisions measured at the LHC, by the ATLAS, CMS, and ALICE Collaborations \cite{Aad:2012gla,CMS:2012qk,Abelev:2012ola}. At RHIC, PHENIX observed similar signals in high multiplicity d-Au collisions \cite{Adare:2013piz}, confirming the findings made at LHC energies. One should note that pA collisions were always considered as baseline and well understood measurements, and the fact that signs of hydrodynamic behavior were observed in such small systems came as a surprise. Understanding this novel behavior of strongly interacting QCD matter is now one of the main threads in high energy nuclear physics.

In this work we perform hydrodynamic simulations of  pA
and  AA collisions at LHC energies. Our goal is to verify whether high multiplicity  pA collisions
are really able to create a quark-gluon plasma near thermodynamic equilibrium. We test our hydrodynamic model by comparing its results to the full set of the experimentally measured transverse momentum two-particle correlation matrix elements. 
We further check if  pA and  AA collisions can be described within the \textit{same} hydrodynamic model.

\section{Hydrodynamic model}

%In this work we simulate high multiplicity  pA collisions and peripheral  AA collisions at LHC energies. 
The initial state of
the collision in our model is calculated using the Monte Carlo Glauber model \cite%
{Miller:2007ri}, extended by including the effects of system's longitudinal anisotropy and of local entropy density fluctuations~\cite{Bozek:2010bi,Bozek:2011if}
as described below. The subsequent dynamics of the system is evolved using
relativistic dissipative hydrodynamics, solved numerically in 3+1 dimensions
with the \textsc{music} approach \cite{Schenke:2010rr}.

The initial entropy density profile at the thermalization time, $\tau
_{0}=0.6$ fm/c, is given by%
\begin{equation}
s\left( \vec{x}_{\bot },\eta ,\tau _{0}\right) =\left( 1\pm \frac{\eta }{y_{%
\mathrm{beam}}}\right) \exp \left[ -\frac{\left( \left\vert \eta \right\vert
-\eta _{0}\right) ^{2}}{2\sigma _{\eta }^{2}} \theta \left(
\left\vert \eta \right\vert -\eta _{0}\right) \right] \times \sum_{i=1}^{N_{\mathrm{%
part}}}\frac{S_{i}}{2\pi \sigma ^{2}}\exp \left( -\frac{\left\vert \vec{x}%
_{\bot }-\vec{x}_{\bot }^{\,i}\right\vert ^{2}}{2\sigma ^{2}}\right) ,
\end{equation}%
where $N_{\mathrm{part}}$ is the number of wounded nucleons, $\left( \vec{%
x}_{\bot }^{\,i},\eta ^{\,i}\right) $ is the position of the $i$--th wounded
nucleon in hyperbolic coordinates and $\pm$ corresponds to the sign of the participating nucleon's longitudinal momentum. 
The wounded nucleons are calculated
taking a nucleon-nucleon inelastic cross section of $\sigma _{\rm NN}=67$ mb.
The parameter $\sigma $ specifies the length scale of the entropy density
fluctuations in the transverse plane and is taken to be in the range 
$\sigma =0.4-0.8$ fm.
Recalling multiplicity fluctuations occurring in proton-proton collisions, we use 
a negative binomial distribution (NBD) to describe individual participant's contribution, $S_{i}$, to the total entropy.
To describe the entropy density profile in the longitudinal direction, we use parameters values $\eta _{0}=2.5$, $\sigma _{\eta }=1.4$ and $y_{\mathrm{beam}}=8.58$ (p-Pb at $\sqrt{s_{\rm NN}} = 5.02$ TeV).

In all our calculations the initial transverse velocity profile is assumed to be zero and
the system starts its hydrodynamic evolution in local thermodynamic
equilibrium. The hydrodynamic equations solved correspond to the usual
continuity equation, $\partial _{\mu }T^{\mu \nu }=0$, which describes
energy-momentum conservation, coupled with a version of Israel-Stewart theory 
\cite{Israel:1979wp}, 
\begin{equation}
\tau _{\pi }\Delta _{\alpha \beta }^{\mu \nu }u^{\lambda }\partial _{\lambda
}\pi ^{\alpha \beta }+\pi ^{\mu \nu }=2\eta \sigma ^{\mu \nu }+\frac{4}{3}%
\tau _{\pi }\pi ^{\mu \nu }\partial _{\lambda }u^{\lambda }\text{,}
\label{eq:pi_munu}
\end{equation}%
which describes the time evolution of the shear-stress tensor, $\pi ^{\mu
\nu }$. Above, we introduced the shear tensor, $\sigma ^{\mu \nu }=\Delta
_{\alpha \beta }^{\mu \nu }\partial ^{\alpha }u^{\beta }$, and the double,
symmetric, traceless projection operator $\Delta _{\alpha \beta }^{\mu \nu
}=\left( \Delta _{\alpha }^{\mu }\Delta _{\beta }^{\nu }+\Delta _{\beta
}^{\mu }\Delta _{\alpha }^{\nu }\right) /2-\Delta _{\alpha \beta }\Delta
^{\mu \nu }/3$. In this work, we neglect the effects of bulk viscous
pressure and of net-baryon number diffusion. The equation of state employed in
all our calculations is the parametrization of lattice QCD calculations by
Huovinen and Petreczky \cite{Huovinen:2009yb}. Also, the shear viscosity coefficient is assumed
to be proportional to the entropy density, $\eta /s=0.\mbox{--}0.08$, and the shear
relaxation time is given by $\tau _{\pi }=3\eta /\left( \varepsilon
+P\right) $, with $\varepsilon $ being the energy density and $P$ the
thermodynamic pressure. The freeze-out procedure is implemented via the 
Cooper-Frye formalism \cite{Cooper:1974mv}, with a freeze-out temperature of $T=150$ MeV. We
direct the reader to Refs.~\cite{Kozlov:2014fqa} for further details of our
model and to Ref.~\cite{Bozek:2011if}, \cite{Bzdak:2013zma}, \cite{Werner:2013ipa} for the hydrodynamic simulations of pA collisions utilizing different types of initial conditions.

\section{Comparison with proton-nucleus data}

\label{comparison}

\begin{figure}[th]
\centering
\includegraphics[width=0.42\textwidth]{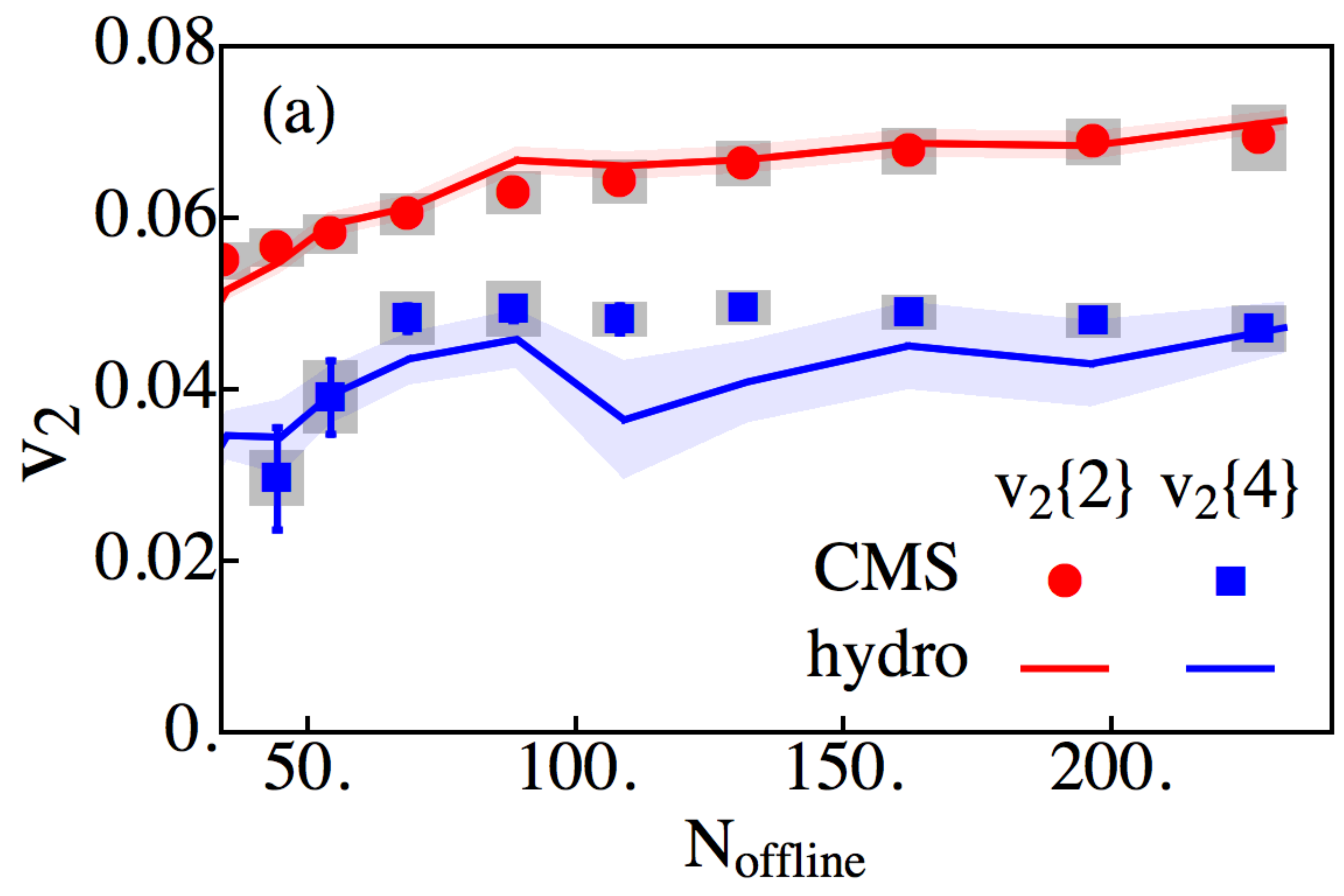}\hspace*{0.8cm} \includegraphics[width=0.42\textwidth]{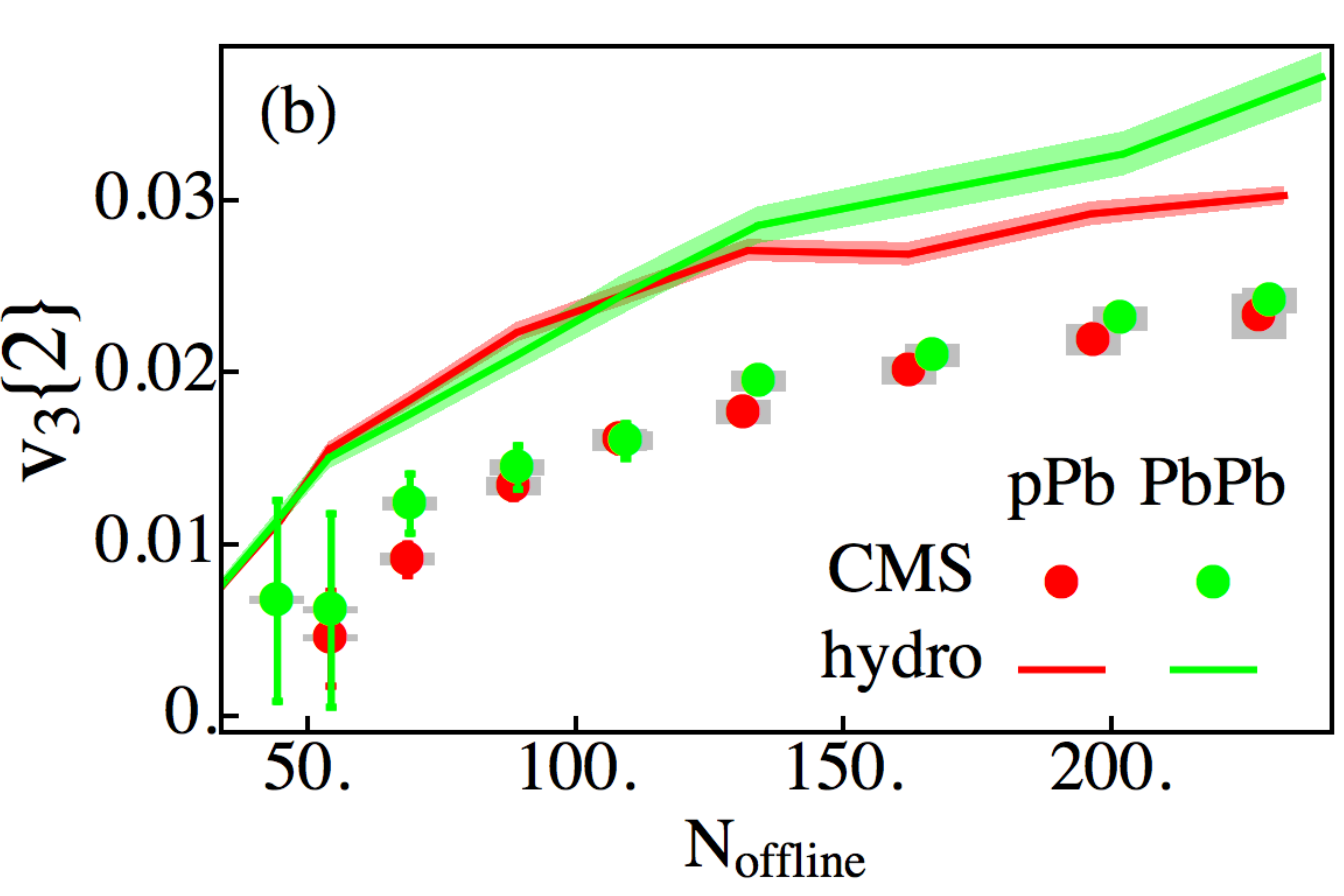}
\caption{Elliptic flow variables compared to calculations from the hydrodynamic approach discussed in the text. The left panel shows CMS data for p-Pb collisions at $\sqrt{s_{\rm NN}} = 5.02$ TeV. The right panel contains p-Pb ($5.02$ TeV) and Pb-Pb ($2.76$ TeV) triangular flow measurements with similar final multiplicities \cite{Chatrchyan:2013nka}, along with results from calculations described in the text with the used parameters values $\sigma=0.4$ fm, $\eta/s=0.08$.}
\label{fig:one}
\end{figure}

We start by comparing our calculations of the integrated two-particle and
four-particle cumulants, $v_{n}\{2\}$ and $v_{n}\{4\}$, to the experimentally
observed values. In Fig.~\ref{fig:one} (a) we show integrated $v_{2}\{2\}$ and $%
v_{2}\{4\}$ as a function of multiplicity for  pA collisions,
with the curves corresponding to the theoretical predictions and the points
to the CMS data \cite{Chatrchyan:2013nka}. We see that
our calculations are in good agreement with the data, supporting the claim
that high multiplicity  pA collisions display collective
behavior. Surprisingly, the agreement with the data remains good even when
the multiplicity is not so large and only visibly starts to break when $N_{%
\mathrm{offline}}<50$. 
We explored features of the hydro calculations, by modifying parameters of granularity ($\sigma$) and viscosity ($\eta/s$) in the value ranges specified above, and verifying, that hydro curves stay in the vicinity of the experimental data. Thus, we expect our hydro model to correctly capture the general experimental trends. To test it, we consider the CMS finding of the remarkably similar magnitude of the $v_3\{2\}$ for p-Pb and Pb-Pb data that, to cite the authors, ``was not trivially expected within a hydrodynamic picture.'' 
%~\cite{Chatrchyan:2013nka} 
However, it naturally follows in our hydro model without any additional fitting, Fig.~\ref{fig:one} (b). 
We have not made efforts to reproduce the absolute magnitude of $v_3\{2\}$, but it appears plausible that the apparent collectivity observed in  pA collisions can have a fluid-dynamical origin.

A more stringent test of the fluid-dynamical nature of  pA
collisions can be obtained from a detailed analysis of the transverse
momentum structure of two-particle correlations \cite{Gardim:2012im}.
Experimentalists measure the full correlation matrix
\begin{equation}
V_{n\Delta }(p_{T}^{a},p_{T}^{b})=\left\langle \frac{1}{N_{\mathrm{pairs}%
}^{a,b}}\sum_{\mathrm{pairs}\{a,b\}}\cos n\Delta \phi \right\rangle ,
\end{equation}%
which includes the regularly discussed two-particle cumulant --- 
with $N_{\mathrm{pairs}}^{a,b}$ being the number of pairs with transverse
momenta $p_{T}^{a}$ and $p_{T}^{b}$ in a given event, $\sum_{\mathrm{pairs}%
\{a,b\}}$ is a summation over this set of pairs, and $\Delta \phi =\phi
^{a}-\phi ^{b}$ their relative azimuthal angle. The brackets denote an
average over events. Note that the two-particle cumulant shown in~Fig.~\ref{fig:one} is a special case of
this general correlator $v_{n}\{2\}=\sqrt{V_{n\Delta }(\bar{p}_{T},
\bar{p}_{T})}$ with $\bar{p}_{T}\in \left[ 0.3, 3.0 \right] \mathrm{GeV/c}$. 
While $v_n$~(Fig.~\ref{fig:one}) probes the overall
magnitude of the correlation, considering comparison of the full matrix (and its elements' ratios) calculated in a hydro model to the experimentally measured one allows to better
study the momentum structure of the correlations. 

It is more convenient and intuitive to perform this comparison after
changing variables from $V_{n\Delta }(p_{T}^{a},p_{T}^{b})$ to $%
r_{n}(p_{T}^{a},p_{T}^{b})\equiv V_{n\Delta }(p_{T}^{a},p_{T}^{b})/\sqrt{%
V_{n\Delta }(p_{T}^{a},p_{T}^{a})V_{n\Delta }(p_{T}^{b},p_{T}^{b})}$, because 
$r_n$ ratio is being bounded by $\pm1$ in fluid-dynamical simulations \cite{Gardim:2012im}. 
As well, the fact that these bounds were satisfied~\cite{CMS:2014vca} is non-trivial, since
otherwise it would imply the failure of the hydrodynamic picture, regardless of the initial condition
or set of parameters employed. 
%{\color{red}{Furthermore, as the integrated $v_n$ probes the overall magnitude of the correlation, the ratio that defines $r_n$ allows to better study the momentum structure of the correlation matrix. }} 

\begin{figure}[th]
\centering
\includegraphics[width=0.66\textwidth]{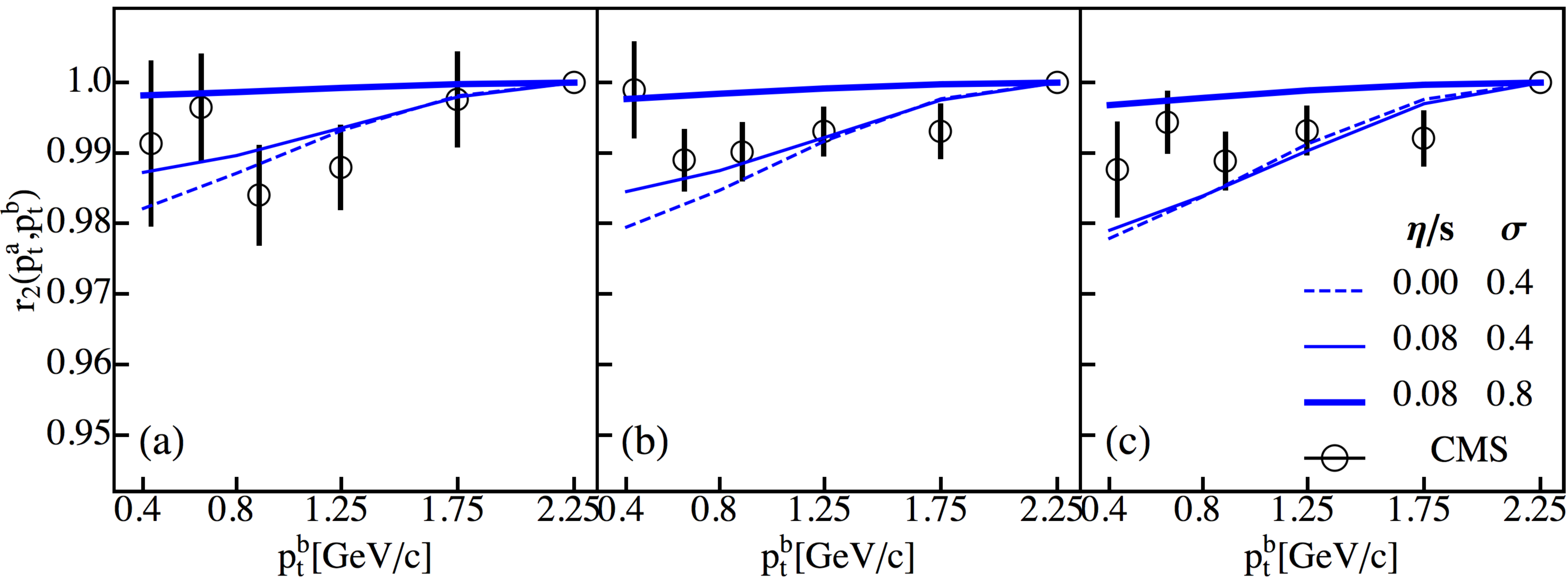}\hspace*{0.8cm} \includegraphics[width=0.2579\textwidth]{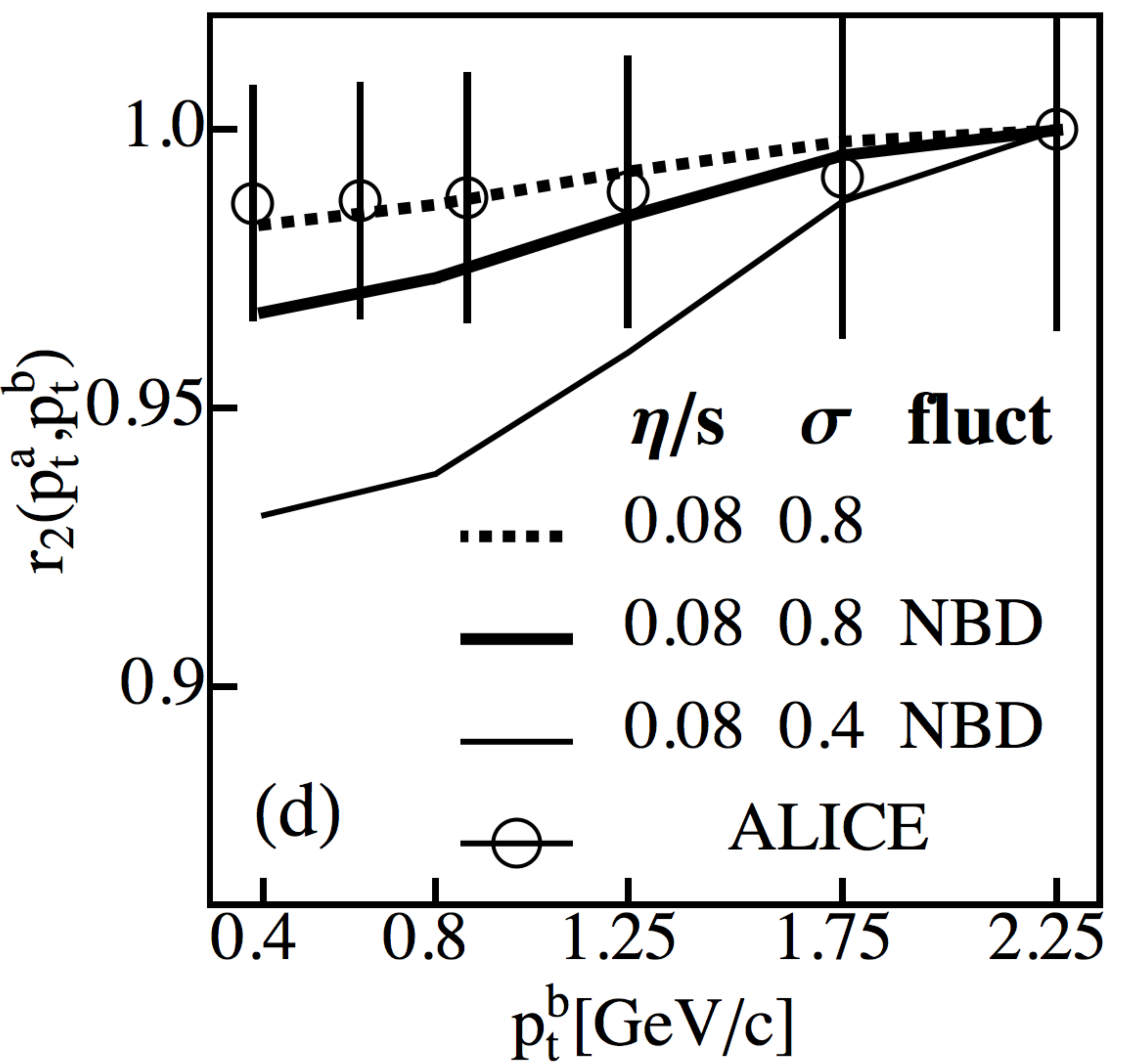}
%\subfigure[]{\includegraphics[width=0.3\textwidth]{r2mult.pdf}} 
%\hspace{1.0cm}}
%\subfigure[]{\includegraphics[width=0.3\textwidth]{r2param.pdf}} 
%\hspace{1.0cm}}
%\subfigure[]{\includegraphics[width=0.3\textwidth]{r2AA.pdf}}
\caption{The pair correlation variable $r_n (p_T^a, p_T^b)$ (only $n=2$ case is shown), for a range of the trigger momentum $p_T^a\in\left[ 2,2.5\right] $ GeV/c, as a function of the associate momentum $p_T^b$. {\it Panels (a), (b), and (c)}: Panel (a) is for a multiplicity class of $220 \leq N_{\rm offline} < 260$, panel (b) is for  $185 \leq N_{\rm offline} < 220$, and panel (c) is for $150 \leq N_{\rm offline} < 185$. The legend in panel (c)  applies also to (a) and (b). The data is by CMS Collaboration \cite{CMS:2014vca}, and the curves are {\it predictions} \cite{Kozlov:2014fqa} of the model described in the text (see also the model predictions for ALICE measurements \cite{Zhou:2014bba}). {\it Panel (d)}: $r_2 (p_T^a, p_T^b)$ for Pb-Pb collisions as measured by the ALICE Collaboration \cite{Aamodt:2011by}. Details are given in the text.}
\label{fig:rn}
\end{figure}

In Fig.~\ref{fig:rn} (a-c), we show $r_{2}(p_{T}^{a},p_{T}^{b})$ for
 pA collisions as a function of $p_{T}^{b}$, for $p_{T}^{a}\in %
\left[ 2,2.5\right] $ GeV/c and for several multiplicity ranges. We have verified that  our results imply a monotonic dependence of $r_{n}$ on the multiplicity, i.e., $r_{n}$  values decrease when we go to lower multiplicities.  In order to probe the sensitivity of this
observable to the granularity and viscosity of this system, we also show
results in the ideal fluid limit, $\eta /s=0$, and with smaller
granularity, $\sigma=0.8$ fm. We observe that the effect of viscosity on $%
r_{n}$ is not very large, but that the effect of granularity is rather
significant. The CMS data clearly favors the
calculation with larger granularity, i.e., $\sigma=0.4$ fm. The fact that $%
r_{n}$ exhibits this sensitivity to the granularity is beneficial, as it could be used
as a probe of the initial condition's granularity in the same way flow is
used to probe the viscous properties of the QCD matter.

In Fig.~\ref{fig:rn} (d), we show $r_{2}(p_{T}^{a},p_{T}^{b})$ for peripheral
 AA collisions as a function of $p_{T}^{b}$, for the same range
in $p_{T}^{a}$ described above, compared to the ALICE data \cite{Aamodt:2011by}. Here, we test the effect of granularity, by
changing $\sigma =0.4$ fm to $\sigma=0.8$ fm, and of multiplicity
fluctuations, by removing the NBD fluctuation of the entropy
produced by each wounded nucleon. While in  AA collisions both
these quantities have a very small effect on the usual flow observables, we
see that they have a considerable effect on $r_{n}$. Surprisingly, the
calculation with larger granularity and with multiplicity fluctuations, which
could reasonably describe the CMS pA collision data, is not in good
agreement with the ALICE AA data. It is puzzling that pA and AA data cannot be
described with the same set of parameters, since one expects that the same
type of fluid is created in both collisions. We will further investigate
this in future work.

\section{Conclusions}

\label{conclusions}

In this work we have shown that hydrodynamic models are able to reasonably
describe a wide range of 
flow variables 
and studied behavior of the $r_n$ 
%collective flow 
observable. This quantity provides a general test of the applicability of hydrodynamics 
(independent of the used parameters and the initial conditions) to the description of high multiplicity pA collisions. 
We have shown that $r_{n}\left( p_{T}^{a},p_{T}^{b}\right) $ is sensitive to local entropy
fluctuations and to the initial granularity of the fluid: it is thus able to
probe aspects of the initial state of the collision that traditional flow
observables cannot. Importantly, the fact that our prediction of the pair correlation observables $r_2$ and $r_3$ were subsequently confirmed by the CMS collaboration measurements adds credence to the line of reasoning presented here. 
Finally, our attempt to describe $r_n$ data for various HIC systems reveals that the simultaneous interpretation of the available 
 AA and  pA collision data within a unified hydrodynamical approach still currently appears challenging.

\textbf{Acknowledgments:} The authors thank P.~Bozek for fruitful discussions. This work
was supported in part by the Natural Sciences and Engineering Research
Council of Canada, and in part by the 
Director, Office of Energy Research, Office of High
Energy and Nuclear Physics, Division of Nuclear Physics,
of the U.S. Department of Energy under Contract No.
DE-AC02-05CH11231. I.K. acknowledges support from the Canadian Institute of Nuclear Physics,  G.S.D. acknowledges support through a Banting Fellowship of the
Natural Sciences and Engineering Research Council of Canada.%, and C.~G.~
%acknowledges support from the Hessian Initiative for Excellence (LOEWE)
%through the Helmholtz International Center for FAIR (HIC for FAIR).


\begin{thebibliography}{99}
%\bibitem{Ackermann:2000tr} K.~H.~Ackermann \textit{et al.} [STAR
%Collaboration], 
%``Elliptic flow in Au + Au collisions at (S(NN))**(1/2) = 130 GeV,''
%Phys.\ Rev.\ Lett.\ \textbf{86}, 402 (2001) [nucl-ex/0009011]. 
%%CITATION = NUCL-EX/0009011;%%
%544 citations counted in INSPIRE as of 15 Jul 2014

%\cite{CMS:2012qk}


%\cite{Aad:2012gla}
\bibitem{Aad:2012gla} G.~Aad, et al., ATLAS Collaboration, 
%``Observation of Associated Near-side and Away-side Long-range Correlations in $\sqrt{s_{NN}}$=5.02 TeV Proton-lead Collisions with the ATLAS Detector,''
Phys.\ Rev.\ Lett.\ {110} (2013) 182302, arXiv:1212.5198 [hep-ex]. 
%%CITATION = ARXIV:1212.5198;%%
%93 citations counted in INSPIRE as of 14 May 2014

\bibitem{CMS:2012qk} S.~Chatrchyan, et al., CMS Collaboration, 
%``Observation of long-range near-side angular correlations in proton-lead collisions at the LHC,''
Phys.\ Lett.\ B {718} (2013) 795, arXiv:1210.5482 [nucl-ex]. 
%%CITATION = ARXIV:1210.5482;%%
%103 citations counted in INSPIRE as of 14 May 2014

%\cite{Abelev:2012ola}
\bibitem{Abelev:2012ola} B.~Abelev, et al., ALICE Collaboration, 
%``Long-range angular correlations on the near and away side in $p$-Pb collisions at $\sqrt{s_{NN}}=5.02$ TeV,''
Phys.\ Lett.\ B {719} (2013) 29, arXiv:1212.2001 [nucl-ex]. 
%%CITATION = ARXIV:1212.2001;%%
%117 citations counted in INSPIRE as of 14 May 2014

%\cite{Adare:2013piz}
\bibitem{Adare:2013piz} 
  A.~Adare, { et al.},  PHENIX Collaboration,
%  %``Quadrupole Anisotropy in Dihadron Azimuthal Correlations in Central $d$$+$Au Collisions at $\sqrt{s_{_{NN}}}$=200 GeV,''
  Phys.\ Rev.\ Lett.\  {111} (2013) 212301,
  arXiv:1303.1794 [nucl-ex].
%  %%CITATION = ARXIV:1303.1794;%%
%  %50 citations counted in INSPIRE as of 14 May 2014

%\cite{Miller:2007ri}
\bibitem{Miller:2007ri} M.~L.~Miller, K.~Reygers, S.~J.~Sanders and
P.~Steinberg, %``Glauber modeling in high energy nuclear collisions,''
Ann.\ Rev.\ Nucl.\ Part.\ Sci.\ {57} (2007) 205, arXiv:nucl-ex/0701025. 
%%CITATION = NUCL-EX/0701025;%%
%485 citations counted in INSPIRE as of 15 Jul 2014

\bibitem{Bozek:2010bi} P.~Bozek and I.~Wyskiel, 
%``Directed flow in ultrarelativistic heavy-ion collisions,''
Phys.\ Rev.\ C {81} (2010) 054902, arXiv:1002.4999 [nucl-th]. 
%%CITATION = ARXIV:1002.4999;%%
%37 citations counted in INSPIRE as of 14 May 2014

%\cite{Bozek:2011if}
\bibitem{Bozek:2011if} 
  P.~Bozek,
  %``Collective flow in p-Pb and d-Pd collisions at TeV energies,''
  Phys.\ Rev.\ C {\bf 85} (2012) 014911,
  arXiv:1112.0915 [hep-ph].
  %%CITATION = ARXIV:1112.0915;%%
  %80 citations counted in INSPIRE as of 02 Aug 2014

%\cite{Schenke:2010rr}
\bibitem{Schenke:2010rr} B.~Schenke, S.~Jeon and C.~Gale, 
%``Elliptic and triangular flow in event-by-event (3+1)D viscous hydrodynamics,''
Phys.\ Rev.\ Lett.\ {106} (2011) 042301, arXiv:1009.3244 [hep-ph]. 
%%CITATION = ARXIV:1009.3244;%%
%250 citations counted in INSPIRE as of 15 Jul 2014


%\cite{Israel:1979wp}
\bibitem{Israel:1979wp} 
  W.~Israel and J.~M.~Stewart,
  %``Transient relativistic thermodynamics and kinetic theory,''
  Annals Phys.\  {118} (1979) 341.
  %%CITATION = APNYA,118,341;%%
  %560 citations counted in INSPIRE as of 01 Aug 2014

\bibitem{Huovinen:2009yb} P.~Huovinen and P.~Petreczky, 
%``QCD Equation of State and Hadron Resonance Gas,''
Nucl.\ Phys.\ A {837} (2010) 26.
%%CITATION = ARXIV:0912.2541;%%

\bibitem{Cooper:1974mv} F.~Cooper and G.~Frye, 
%``Comment on the Single Particle Distribution in the Hydrodynamic and Statistical Thermodynamic Models of Multiparticle Production,''
Phys.\ Rev.\ D {10} (1974) 186. %%CITATION = PHRVA,D10,186;%%


%\cite{Kozlov:2014fqa} 
\bibitem{Kozlov:2014fqa} I.~Kozlov, M.~Luzum, G.~Denicol, S.~Jeon and
C.~Gale, 
%``Transverse momentum structure of pair correlations as a signature of collective behavior in small collision systems,''
arXiv:1405.3976 [nucl-th]. %%CITATION = ARXIV:1405.3976;%%
%2 citations counted in INSPIRE as of 15 Jul 2014

%\cite{Bzdak:2013zma}
\bibitem{Bzdak:2013zma} 
  A.~Bzdak, B.~Schenke, P.~Tribedy and R.~Venugopalan,
  %``Initial state geometry and the role of hydrodynamics in proton-proton, proton-nucleus and deuteron-nucleus collisions,''
  Phys.\ Rev.\ C {87} (6) (2013) 064906, 
  arXiv:1304.3403 [nucl-th].
  %%CITATION = ARXIV:1304.3403;%%
  %82 citations counted in INSPIRE as of 10 Sep 2014

%\cite{Werner:2013ipa}
\bibitem{Werner:2013ipa} 
  K.~Werner, M.~Bleicher, B.~Guiot, I.~Karpenko and T.~Pierog,
  %``Evidence for flow in pPb collisions at 5 TeV from v2 mass splitting,''
  Phys.\ Rev.\ Lett.\  {112} (2014) 232301,
  arXiv:1307.4379 [nucl-th].
  %%CITATION = ARXIV:1307.4379;%%
  %36 citations counted in INSPIRE as of 10 Sep 2014

%\cite{Chatrchyan:2013nka}
\bibitem{Chatrchyan:2013nka} S.~Chatrchyan, et al., CMS
Collaboration, 
%``Multiplicity and transverse momentum dependence of two- and four-particle correlations in pPb and PbPb collisions,''
Phys.\ Lett.\ B {724} (2013) 213, arXiv:1305.0609 [nucl-ex];

Additional data on the CMS public wiki: 
\verb|https://twiki.cern.ch/twiki/bin/view/CMSPublic/PhysicsResultsHIN13002|.
%%CITATION = ARXIV:1305.0609;%%
%55 citations counted in INSPIRE as of 14 May 2014

%%\cite{Chatrchyan:2013eya}
%\bibitem{Chatrchyan:2013eya} S.~Chatrchyan \textit{et al.} [CMS
%Collaboration], 
%%``Study of the production of charged pions, kaons, and protons in $p$Pb collisions at $\sqrt{s_{NN}} = 5.02$ TeV,''
%arXiv:1307.3442 [hep-ex]. %%CITATION = ARXIV:1307.3442;%%
%%27 citations counted in INSPIRE as of 15 May 2014

%%\cite{AlexanderMilovonbehalfoftheATLAS:2014rta}
%\bibitem{AlexanderMilovonbehalfoftheATLAS:2014rta} A.~Milov [ATLAS
%Collaboration], 
%``Particle production and long-range correlations in p+Pb collisions with the ATLAS detector,''
%arXiv:1403.5738 [nucl-ex]. %%CITATION = ARXIV:1403.5738;%%


%\cite{Gardim:2012im}
\bibitem{Gardim:2012im} F.~G.~Gardim, F.~Grassi, M.~Luzum and
J.-Y.~Ollitrault, 
%``Breaking of factorization of two-particle correlations in hydrodynamics,''
Phys.\ Rev.\ C {87} (3) (2013) 031901, arXiv:1211.0989
[nucl-th]. %%CITATION = ARXIV:1211.0989;%%
%17 citations counted in INSPIRE as of 14 May 2014

%\cite{CMS:2014vca}
\bibitem{CMS:2014vca} CMS Collaboration, 
%``Factorization breakdown of two-particle correlations in pPb and PbPb collisions at CMS,''
CMS-PAS-HIN-14-012;%%CITATION = CMS-PAS-HIN-14-012;%%

D. Devetak, { et al}., CMS Collaboration, these proceedings.

%\cite{Zhou:2014bba}
\bibitem{Zhou:2014bba} Y.~Zhou, for the ALICE Collaboration, 
%``Searches for $p_{\rm T}$ dependent fluctuations of flow angle and magnitude in Pb--Pb and p--Pb collisions,''
arXiv:1407.7677 [nucl-ex]. %%CITATION = ARXIV:1407.7677;%%

%\cite{Aamodt:2011by}
\bibitem{Aamodt:2011by}
  K.~Aamodt, { et al.},  ALICE Collaboration,
  %``Harmonic decomposition of two-particle angular correlations in Pb-Pb collisions at $\sqrt{s_{NN}}=2.76$ TeV,''
  Phys.\ Lett.\ B {708} (2012) 249,
  arXiv:1109.2501 [nucl-ex].
  %%CITATION = ARXIV:1109.2501;%%
  %113 citations counted in INSPIRE as of 02 Aug 2014

\end{thebibliography}
\end{document}